\documentstyle[draft,aps,epsfig]{revtex}
\begin{document}
\draft
\tightenlines
\title{Higher Resonance Contamination of $\pi NN$ Couplings Obtained \\
Via the Three-Point Function Method in QCD Sum Rules}
\author{Kim Maltman} 
\address{Department of Mathematics and Statistics, York University, \\
          4700 Keele St., North York, Ontario, CANADA M3J 1P3 \\ and}
\address{Special Research Center for the Subatomic Structure of Matter, \\
          University of Adelaide, Australia 5005}
\maketitle
\begin{abstract}
We investigate the size of potential higher pseudoscalar resonance
contaminations of the estimates of isospin-conserving and
isospin-violating $\pi NN$ couplings obtained using the 3-point
function method in QCD sum rules.  For the isospin-conserving case
it is shown that conventional models of the isovector pseudoscalar
spectral function imply resonance decay constants large enough
to create significant contaminations, and that assuming
these models are incorrect, and that the
decay constants are actually much smaller, implies physically implausible
values for the flavor-breaking quark condensate ratios.  For the
isospin-violating case it is shown explicitly that such resonance
contamination is unavoidably present and precludes using the 3-point
function method as a means of estimating the at present
unmeasured isospin-violating $\pi NN$ couplings.
\end{abstract}
\pacs{13.75.Gx,11.30.Hv,11.55.Hx,24.85.+p}
\section{Introduction}
The general framework of QCD sum rules\cite{svz,rry,narison} 
has recently proven 
popular as an approach to the computation of observable of
relevance to problems in nuclear and few-body physics which has,
in contrast to many effective hadronic models, a rather direct
connection with QCD itself.  The method is attractive, first, because
it is based on rather general properties of the underlying
field theory (the operator product expansion (OPE), analyticity,
unitarity, and the existence of appropriately subtracted dispersion
relations) and, second, because it provides a means of relating
integrals over physical spectral densities to the behavior
of correlators at large spacelike momenta (obtained via the OPE)
which, via the Borel transformation of the original dispersion
relation, simultaneously exponentially suppresses higher energy
contributions to the physical spectral integrals and factorially
suppresses the contributions to the OPE associated with higher
dimension operators.  This means that one has the quite
reasonable hope of, in favorable circumstances,
constructing a sum rule relating observable parameters
(such as masses, couplings and decay constants) occuring on
the phenomenological side of the sum rule to a small number
of vacuum condensates (which parametrize non-perturbative
effects in QCD) appearing on the OPE side of the sum rule.

In the case of 2-point functions, it is rather easy to see what
``circumstances'' are favorable to such an analysis.  First, one
should know that one is considering a channel where the relevant
spectral function consists of a lowest lying single resonance
contribution well-separated from higher resonance and/or 
continuum pieces.  This allows one to choose a Borel mass
to strongly suppress the contributions to the weighted spectral
integral from 
the more complicated part of the physical
spectral function, and hence obtain an expression for
the phenomenological side of the sum rule that is dominated
by a few observable parameters associated with the lowest lying
physical state.  Second, for such Borel masses, it must 
simultaneously be the case that the OPE side of the sum rule
is well-converged at operators of low enough dimension that 
the corresponding vacuum values are already known from other
analyses.  Leinweber\cite{sceptics} has provided a very clear discussion
of the criteria to be satisfied for the applicability of 
the sum rule method, and also a procedure for sensibly
checking the validity of these criteria in a given case.

In the case of observables like the isospin-conserving and 
isospin-breaking $\pi NN$ couplings, the situation is slightly more
complicated.  In the past, two different approaches have been taken
to estimating these couplings in the QCD sum rule framework.

The first of these approaches is the 3-point function 
method\cite{rry,rry83,reinders84,mh97}.  
In this approach
one considers the 3-point vacuum correlator
\begin{equation}
G_{\pi^a NN} (p_1,p_2,q) = 
\int d^4 x_1 d^4 x_2 \exp \left({ip_1 x_1}-{ ip_2 x_2}\right)
\left \langle 0 \vert {\rm T}\left( 
\chi_N (x_1) P_{I=1}^a(0) {\bar{\chi_N}} (x_2)\right) \vert 0 \right \rangle ,
\label{3pt}
\end{equation}
where $a$ labels the isospin of the pion ($\pm$ or $0$), $N$ 
stands for either proton or neutron, $P_{I=1}^a\equiv P_1^a$ is a pseudoscalar
interpolating field for the pion, and $\chi_N$ is the Ioffe
interpolating field for the nucleon.  In what follows we will
use the following notation for the pseudoscalar currents:
\begin{eqnarray}
P_f &=& \bar{q}_f  i\gamma_5 q_f \nonumber \\
P^+ &=& \sqrt{2} \bar{u}i \gamma_5  d  ,
\label{psdef}
\end{eqnarray}
where $f$ is a flavor label $f=u,d,s$.  The $a=0$ component of
the isovector current multiplet is, as usual,
\begin{equation}
P^0_1 = \bar{u} i\gamma_5 u\ -\ \bar{d} i\gamma_5 d\ .
\label{ps10}
\end{equation}
In what follows we will also require the flavor neutral
isoscalar currents.  For these we can use either the strange
and light quark combinations 
\begin{eqnarray}
P_0^\ell &=& \bar{u} i\gamma_5 u\ +\ \bar{d} i\gamma_5 d\
\nonumber \\
P_0^s &=& \bar{s} i \gamma_5 s
\label{i0mass}
\end{eqnarray}
or the singlet and octet combinations
\begin{eqnarray}
P_0^0 &=&  \left[ \bar{u} i\gamma_5 u\ +\ \bar{d} i\gamma_5 d\
\ +\ \bar{s} i \gamma_5 s\right] /\sqrt{3} \nonumber \\
P_0^8 &=& \left[  \bar{u} i\gamma_5 u\ +\ \bar{d} i\gamma_5 d\
\ -\ 2\, \bar{s} i \gamma_5 s \right]/\sqrt{3}
\label{i0flavor}
\end{eqnarray}
as a basis.
The Ioffe currents for the nucleons are defined by
\begin{eqnarray}
\chi_p &=& \epsilon_{abc} \left [ \left ( [u^a]^T {\cal C} \gamma_\mu
u^b \right ) \gamma_5 \gamma^\mu d^c \right ] , \\
\chi_n  &=& \epsilon_{abc} \left [ \left ( [d^a]^T {\cal C} \gamma_\mu
d^b \right ) \gamma_5 \gamma^\mu u^c \right ] .
\label{nioffe}
\end{eqnarray}
Schematically, one then analyses the correlator using both a hadronic
model (which involves the relevant $\pi NN$ couplings as parameters)
and the OPE, and matches the two representations, appropriately
transformed, to extract the $\pi NN$ couplings.  In order to
perform this analysis, it is necessary that the momentum $q$
on the external pion leg of the correlator be large and spacelike,
otherwise the OPE of the correlator will not be valid truncated
to the low-dimension operators for which the analysis is practical.
This condition, however, means that one is rather far from the
pion pole.
This problem is dealt with by looking for terms in the OPE which
have the same Lorentz structure as the pion contribution and
in addition have a pole of the form $1/q^2$.  However, as stressed
by Birse and Krippa\cite{bk96}, this proceedure is inherently rather
dangerous, since it is not {\it a priori} possible to separate
contributions from the pion and from higher resonances to the
coefficient of $1/q^2$.  Since, moreover, one must certainly
work at $Q^2 \equiv -q^2 > 1$ GeV$^2$, i.e., rather
far from the pion pole, one may no longer reasonably count
on the proximity of this pole to conclude that the pion
contribution is dominant, as it would be for small $Q^2$.
The reliability of the 3-point function method thus rests crucially
on the assumption that the higher resonance contributions, in the
region of $Q^2$ values under consideration, are small.  The 
plausibility of this assumption has not previously, to our knowledge,
been investigated.

The second approach to the $\pi NN$ couplings is the 2-point function
method\cite{rry,reinders84,bk96,sh95}.  In this approach, one
considers the 2-point correlator
\begin{equation}
\Pi (p,q) = i\int \, d^4x \, \exp (ipx) \langle 0\vert
{\rm T}\left( \chi_N (x) \bar{\chi}_N(0)\right)\vert \pi^a(q)\rangle\ .
\label{2pt}
\end{equation}
For large spacelike values of $p$, the OPE for the 
product of the two nucleon interpolating
fields can presumably be truncated at operators of relatively low
dimension.  One may then look at the vacuum-to-pion matrix elements
of these operators in order to estimate the $\pi NN$ couplings.
As has been stressed by many authors,
in order to be able to remove contributions associated with
$N\rightarrow N^*$ transitions, one should look not at the
$\gamma_5$ term in the OPE, but rather the structure $q\llap/
\gamma_5$\cite{bk96,bk83,is84,ioffe1,ioffe2}.
The advantage of the 2-point method is, of course, that it
completely avoids the problem of potential contamination from
higher resonance contributions, which is unavoidable in the
3-point function method.  The disadvantage of the method is
that, although the vacuum-to-pion matrix elements of the
lowest dimension operators can all be evaluated with good
accuracy using chiral perturbation theory (ChPT)\cite{gl85},
those for the higher dimension operators are less certain.
Thus, for example, the uncertainty in the value of the
matrix element
\begin{equation}
g_s\langle 0\vert \bar{d}\, \tilde{G}^{\mu\nu}\gamma_\nu u\vert
\pi^+(q)\rangle
\end{equation}
leads to uncertainties of $\sim 15\%$ in the prediction for
the isospin-conserving $\pi NN$ coupling using the 2-point
function method\cite{bk96}.

To date, the isospin-conserving $\pi NN$ coupling has been
considered using both the 2-point\cite{rry,reinders84,bk96,sh95}
and 3-point\cite{rry,rry83,reinders84} 
function methods,
and the isospin-violating $\pi NN$ couplings using the
3-point function method\cite{mh97}.  The isospin-conserving coupling
is, of course, known experimentally, so the point of computing it
using QCD sum rules is primarily to test the plausibility of
the assumptions and truncations that go into the evaluation.
The hope is that success in computing the coupling in the
isospin-conserving case might serve as an indication of the
reliability of the approach employed and hence make the analogous
calculation of the unmeasured isospin-violating couplings
also plausibly reliable.  Since both the 2-point and
3-point function treatments are successful in this regard, 
albeit it with significant
theoretical errors, it would seem reasonable to attempt to
proceed to the isospin-violating case using either method.
A first attempt in this direction was made by Meissner and
Henley\cite{mh97}, employing the 3-point function method.
Since the isospin violating 
coupling has not yet been experimentally determined,
this estimate of is considerable potential interest, particularly
in view of the recent revival of interest in the question of isospin-breaking
in few-body systems (see for example the review of Ref.~
\cite{mnsrev} for an extensive discussion of the situation up to
1990, and Ref.~\cite{kmcw} for a list of more recent papers on 
the subject).  
We would, therefore, like to understand whether, given the
potential problems of the 3-point function method, this estimate
is reliable or not.

In the present paper, therefore,
we will investigate the question
of higher resonance contaminations in the 3-point function
method, which is the biggest potential roadblock to
using the method in the isospin-violating case.  We will
show that existing (albeit model-dependent) understanding
of the spectral density in the isovector pseudoscalar
channel implies that significant contamination is present,
even in the isospin-conserving case, and that requiring this
understanding to be incorrect, and the contamination to be
small, is equivalent to rather strong (and physically
implausible) statements about
the values of flavor-breaking ratios of quark condensates.
We will then proceed to show that certain features of
the isospin-violating analysis itself also clearly indicate
the presence of significant higher resonance contamination,
implying that the 3-point function method cannot be reliably
employed to extract the isospin-violating $\pi NN$ couplings.

\section{The Isospin-Conserving Analysis}

One does not know, {\it a priori}, the size of the couplings
of the excited isovector pseudoscalar mesons, $\pi (1300)$,
$\pi (1800)$, $\cdots$ to the nucleon.  Thus, in order to be
certain that the 3-point function method for 
extracting $g_{\pi NN}$ is not contaminated
by contributions from these resonances it is necessary that
the conditions
\begin{equation}
\frac{f_{\pi^\prime}m^2_{\pi^\prime}}{(Q^2+m^2_{\pi^\prime})}<<
\frac{f_{\pi}m^2_{\pi}}{(Q^2+m^2_{\pi})}
\end{equation}
be satisfied, where $f_M$ is the decay constant for meson $M$
and $\pi^\prime$ stands for any of the excited $I=1$
pseudoscalar mesons.  

The excited state pseudoscalar decay constants 
are not known experimentally and,
owing to the fact that they have a chiral suppression,\cite{svz}
will be very difficult to measure.  (The $\pi (1300)$ decay
constant could, in principle, be measured by separating the small
pseudoscalar contribution from the large overlapping $a_1$
contribution in $\tau$ decays via a detailed spin-parity 
analysis\cite{expidea}.)  However, they {\it are} related to
the light current quark mass combination $m_u+m_d$ via a series
of finite energy sum rules\cite{bpr}.  The most recent analysis
of this mass combination\cite{bpr} models the continuum part of
the spectral function in the isovector pseudoscalar channel
in terms of a sum of $\pi (1300)$ and $\pi (1800)$ resonances.
The relative contribution of the two resonances in the 
model continuum spectral function is constrained by
duality, and the overall normalization is set by assuming the
threshold value of the continuum spectral function, which can
be estimated using ChPT at tree level, is saturated by the tails
of the resonance contributions.  If we assume the model
spectral functions so constructed (which are tied to the usually
quoted values of the $\overline{MS}$ masses) are reasonable, then
we can read off the corresponding values of the decay constants
for the $\pi (1300)$ and $\pi (1800)$.  Taking the model with the
best duality fit from Ref.~\cite{bpr}, we find
\begin{eqnarray}
f_{\pi (1300)}&=&2.2\ {\rm MeV} \nonumber \\
f_{\pi (1800)}&=&1.0\ {\rm MeV}.
\label{decayconstants}
\end{eqnarray}
Forming the products of the couplings, $f_M m^2_M$, to the pseudoscalar
$I=1$ current, and the propagators, $1/(Q^2+m^2_M)$, evaluated at
$Q^2\sim 1\ {\rm GeV}^2$, we then find
\begin{equation}
\left[ \frac{f_M m^2_M}{(Q^2+m^2_M)}\right]_{Q^2=1\ {\rm GeV}^2}=
1.8,\ 2.7,\ {\rm and\ } 0.76\ {\rm MeV}
\end{equation}
for $M=\pi ,\ \pi (1300),\ {\rm and}\ \pi (1800)$, respectively.
For such values of the excited pseudoscalar decay constants, therefore, one
would be forced to conclude that the contamination from, certainly
the $\pi (1300)$, and most likely also the $\pi (1800)$, would be far
too large to make the method reliable.

Of course, one might object that the above argument, relying
as it does on the model spectral functions of Ref.~\cite{bpr},
is model-dependent and therefore not conclusive.  Indeed,
the validity of the method of Ref.~\cite{bpr} for
setting the overall normalization (and
hence the overall scale of the corresponding decay constants in
Eq.~(\ref{decayconstants})) has been questioned\cite{bgm97},
leading to the suggestion that the normalization might actually be
significantly smaller than employed in Ref.~\cite{bpr}.  
However, even if the normalization of
the continuum spectral function were to be decreased by {\it an order
of magnitude,} the corresponding decay constants would be decreased
only by a factor $\sim 3$, leaving the product of the $\pi (1300)$
coupling to the current and propagator at the level of $\sim 50\%$
of that of the $\pi$.

It thus appears extemely unlikely that the couplings of the
higher pseudoscalar resonances can be neglected in the 3-point
function analysis of $g_{\pi NN}$.  Since, however, one does not
actually have an experimental value for $f_{\pi (1300)}$, one 
is still free to imagine
that the decay constants are, for
some reason,
{\it much} more than just an order of magnitude smaller than those
corresponding to the model spectral functions of Ref.~\cite{bpr}
(for example, a spectral function a factor of $60$ times smaller than
that of Ref.~\cite{bpr} would bring the $\pi (1300)$ coupling times
propagator factor to below $20\%$ of the corresponding product
for the $\pi$).  This appears a rather unlikely prospect, but is
one that cannot be presently ruled out.  However, it is important
to realize that such an assumption is not is not without other 
non-trivial consequences.

Let us, therefore, for the moment, accept the (albeit unlikely) prospect of
extremely small excited pseudoscalar decay constants and consider
the consequences of such an assumption for the quark masses and condensates.
The first consequence is obvious, namely, if we strongly suppress the continuum
contribution to the finite energy sum rule analysis for the quark mass,
then we commit ourselves to significantly lower values of $m_u+m_d$,
of order $6$ MeV (in the $\overline{MS}$ scheme, at a scale of
$1$ GeV$^2$, this value to be compared with 
the conventionally quoted value $12$ MeV).  
This is not necessarily a problem, since a recent
analysis of world lattice data also suggests a significantly lowered
value of $m_u+m_d$\cite{bg96}, a possibility also noted
in Ref.~\cite{bgm97}.  It does, however, force one to significantly
larger values of the light quark condensate, which can cause problems
for the stability of the sum rule for the nucleon mass\cite{birse}.  
Moreover, such an assumption actually corresponds to rather strong
constraints on the ChPT low energy constant (LEC), $H_2^r$,
(where we adhere to the notation of Gasser and Leutwyler\cite{gl85}
throughout), which LEC governs the flavor breaking of the quark 
condensates\cite{gl85}.
The reason for the existence of this constraint is that
the inverse weighted (corresponding to $n=-1$
in the notation of Ref.~\cite{bpr}) 
finite energy sum rule for the
correlator
\begin{eqnarray}
\label{eq: two-point funct.}
\Psi_{5}(q^2) \equiv{}&  i\int\,d^4 x e^{iq\cdot x}
  \langle 0\vert T\{\partial^{\mu}A_{\mu}^{(-)}(x),
  \partial^{\nu}A^{(+)}_{\nu}(0)\}\vert 0\rangle ,  \nonumber \\ 
              ={}& (m_{d}+m_{u})^2            i\int\,d^4 x e^{iq\cdot x}
  \langle 0\vert T\{P^{(-)}(x),P^{(+)}(0)\}\vert 0\rangle \ 
\end{eqnarray}
with $A^{(+)}_\mu$ the charged isovector axial current and
$P^{(+)}$ the corresponding charged isovector pseudoscalar current,
can be re-written as a sum rule for the continuum portion of the
pseudoscalar spectral function, as follows:
\begin{equation}
\int_{9m^2_\pi}^s\, \frac{dt}{t} \frac{ {\rm Im}\, \Psi_5(t)}{\pi}
= \frac{3}{8\pi^2}\left( m_u(s)+m_d(s)\right)^2 s\left[
1+R_0(s)\right] - \frac{8m_\pi^4 f_\pi^4}{F^4}\left( 2L_8^r-H_2^r\right)
\label{contsr}
\end{equation}
where $m_{u,d}(s)$ are the $\overline{MS}$ running masses at scale $s$,
$R_0(s)$ contains perturbative corrections\cite{bpr}, $F$ is a
leading order ChPT LEC, equal to the pion decay constant in the chiral
limit, and $2L_8^r-H_2^r$ is a scale-independent combination of
fourth order ChPT LEC's.  If we suppress the integral on the
LHS by a large factor like $60$, and also the running masses by
of order a factor of $2$, it turns out that we drive $2L_8^r-H_2^r$
to values more than $3$ times smaller than those obtained
in Ref.~\cite{bpr}.  
This in turn implies that $H_2^r(m_\eta )$ 
must {\it necessarily} be positive.  We now argue that such values
for $H_2^r(m_\eta )$ lead to physically implausible predictions
for the ratios of quark condensates.
  
To see this, note that, once $H_2^r$ is fixed, 
the flavor breaking ratios of quark condensates are simultaneously
fixed at next-to-leading order in the chiral expansion\cite{gl85},
for example,
\begin{equation}
\frac{\langle 0\vert \bar{s} s\vert 0\rangle}
{\langle 0\vert \bar{u} u\vert 0\rangle}=1+3\mu_\pi -2\mu_K
-\mu_\eta +\frac{8(m_K^2-m_\pi^2 )}{F^2}\left( 2L_8^r +H_2^r\right)\ ,
\label{condratio}
\end{equation}
where $\mu_M =m_M^2\log (m_M^2/\mu^2 )/32\pi^2F^2$, with $\mu$
the ChPT renormalization scale.  With $H_2^r(m_\eta )>0$ and
$L_8^r(m_\eta )=(1.1\pm 0.3)\times 10^{-3}$\cite{bkm95}, we see from
Eq.~(\ref{condratio})
that 
\begin{equation}
\frac{\langle 0\vert \bar{s} s\vert 0\rangle}
{\langle 0\vert \bar{u} u\vert 0\rangle}>1.38\ .
\end{equation}
Thus, assuming that the excited
resonance decay constants are sufficiently small to be able
to neglect their contributions to the 3-point function sum rule
simultaneously commits one to the highly unnatural situation
of a strange quark condensate larger in magnitude than the light quark
condensate.  In addition, one finds that, owing to the
relation between the flavor breaking and isospin breaking
condensate ratios given by Eq.~(9.5) of Ref.~\cite{gl85},
such a value for the strange to up quark condensate ratio implies,
for the isospin-breaking condensate ratio, $\gamma$, defined by
\begin{equation}
\gamma\equiv \frac {\langle 0\vert \bar{d}d\vert 0\rangle
-\langle 0\vert \bar{u}u\vert 0\rangle}{\langle 0\vert \bar{u}u\vert 0\rangle}
\ ,
\label{defngamma}
\end{equation}
a value 
\begin{equation}
\gamma >1.5\times 10^{-3}\ ,
\end{equation}
in contradiction with extractions of $\gamma$
from a variety of sources,\cite{narison,nar87,DdR87,nar89,adi93,ei93} 
all of which obtain $\gamma <0$.  

In view of the results of the last paragraph, we conclude that the
hypothesis that one may neglect the higher resonance contamination
in the 3-point function analysis of the isospin conserving
coupling $g_{\pi NN}$ is a highly
unpalatable one.  In the next section, we will turn to the case
of the analysis of the isospin-breaking couplings and demonstrate
more directly the presence of analogous higher resonance contaminations.

\section{The Isospin-Violating Analysis}

In this section we will concentrate on the 3-point function analysis
of the difference of $\pi^0 pp$ and $\pi^0 nn$ couplings,
\begin{equation}
\delta g = g_{\pi^0 nn}-g_{\pi^0 pp}.
\end{equation}
In order to perform this analysis it is necessary to take into account
the fact that $P_1^0$ is not a suitable interpolating field for
the physical $\pi^0$, if one wishes to
treat isospin breaking effects.  This follows from the observation that
\begin{equation}
\langle 0\vert P_1^0\vert \eta\rangle \not= 0.
\end{equation}
As a consequence, if one were to use $P_1^0$ as
$\pi^0$ interpolating field, then even if one
could ignore higher resonance contributions, the result of the
analysis would be a mixture of the isospin breaking $\pi$ coupling
and the isospin conserving $\eta$ coupling (the latter multiplied
by an isospin breaking factor describing the coupling of the
$\eta$ to the $I=1$, $I_z=0$ current).  As noted above, there is
no means to separate the contributions corresponding to different
mesons in the 3-point function approach.  Meissner and Henley\cite{mh97},
who first performed the isospin violating analysis,
dealt with this problem by choosing a current combination with
no vacuum-to-$\eta$ matrix element, namely,
\begin{equation}
P_{\pi^0}\equiv P_1^0+\epsilon P_0^8,
\label{mhinterp}
\end{equation}
where the pseudoscalar currents are as defined above and the choice
\begin{equation}
\epsilon =\theta_0 =\frac{\sqrt{3}}{4}\frac{m_d-m_u}{m_s-\hat{m}}
\end{equation}
where $\hat{m}=(m_u+m_d)/2$, and $\theta_0$ is the leading
order $\pi$-$\eta$ mixing angle\cite{gl85}, ensures that
\begin{equation}
\langle 0\vert P_{\pi^0}\vert \eta\rangle =0
\end{equation}
to leading order in the chiral expansion.  The choice of interpolating
field with this property is not unique\cite{kmcw}; in fact, for
any $\alpha$, defining
\begin{eqnarray}
P(\alpha )&=&(P_u-P_d)+\epsilon (\alpha )\Bigl[ \alpha (P_u+P_d)+(1+\alpha )P_s
\Bigr]\nonumber\\
&=& (P_u-P_d)+\epsilon (\alpha ){\frac{1}{\sqrt 3}}\Bigl[ -P_8+(3\alpha +1)
P_0\Bigr]\ ,
\label{alphafamily}
\end{eqnarray}
one may find an $\epsilon (\alpha )$ such that
\begin{equation}
< 0|P(\alpha )|\eta >=0\ .
\label{constraint}
\end{equation}
The set of solutions of Eq.~(\ref{constraint}), as a function
of $\alpha$, have been worked out to next-to-leading order in
the chiral expansion in Ref.~\cite{kmcw}.  Among other
results of this analysis, it is found that the next-to-leading
corrections are significant; for example, at next-to-leading
order, the Meissner-Henley field choice must be modified to
\begin{equation}
P^\prime_{\pi^0}\equiv P_1^0+1.27\, \theta_0 P_0^8 \ 
\label{mhinterp1loop}
\end{equation}
if one wishes to maintain zero vacuum-to-$\eta$ matrix element.

We will now explain why the existence of the above family of
potential $\pi^0$ interpolating fields is of relevance to our current
discussion.  Note that the choice of $P_{\pi^0}$ was made by Meissner
and Henley with the problem of potential higher resonance
contamination in mind.  Indeed, they argued that this choice of interpolating
field is the one that would suppress possible $\eta^\prime$ 
contributions\cite{mh97}.  It
turns out, however, that this is not the case, and the reason
that it fails to be so leads us immediately into a consideration
of the larger class of $\pi^0$ interpolating fields.

Let us, therefore, first outline why the interpolating field
$P_{\pi^0}$ necessarily has non-vanishing coupling to the $\eta^\prime$.
In the chiral limit, of course, there is no flavor breaking whatsoever,
hence no isospin breaking couplings, and no flavor breaking meson
decay constants.  Once we introduce the quark mass matrix, with
its flavor breaking $m_s-\hat{m}$ difference and isospin breaking
$m_d-m_u$ difference, all flavor and isospin breaking effects
are potentially present.  In the case of the isospin breaking
and flavor breaking decay
constants of the $\eta^\prime$, one may obtain a leading estimate
of the ratio of these decay constants using $SU(3)_F$ arguments.
Indeed, we know that the breaking is produced by the quark mass
matrix, which has the following decomposition into singlet,
octet isovector and octet hypercharge components
\begin{equation}
M={\frac{1}{3}}(m_s+2\hat{m})
-{\frac{1}{2}}(m_d-m_u)\lambda_3 -{\frac{1}{\sqrt{3}}}(m_s-\hat{m})
\lambda_8\ .
\label{mdecomp}\end{equation}
To leading order in the isospin-breaking and flavor-breaking
mass differences, therefore, the vacuum-to-$\eta^\prime$ matrix elements
of $P_1^0$ and $P_0^8$ are given simply by the product of the mass-dependent
coefficients of $\lambda_3$ and $\lambda_8$ in Eq.~(\ref{mdecomp}) with
a common $8_F\times 8_F\rightarrow 1_F$ reduced matrix element.
Recasting the ratio of mass factors in terms of the mixing
angle $\theta_0$ defined above, one then finds, straightforwardly, that
\begin{equation}
<0|P_{\pi^0}|\eta^\prime >=3\theta_0<0|P_0^8|\eta^\prime >
\end{equation}
where $<0|P_0^8|\eta^\prime >={\cal O}(m_s-\hat{m})$.  The RHS is thus
non-zero, of ${\cal O}(m_d-m_u)$, and in fact has a numerical
enhancement (the factor $3$) brought about by the fact that
the couplings of the $\eta^\prime$ to the $P_1^0$ and $P_0^8$
components of $P_{\pi^0}$ add coherently.  Note that a similar
argument, using a first order treatment
of flavor and isospin breaking, would
predict that the isospin violating and flavor violating
axial current $\eta^\prime$ decay constants were of the same
sign, in agreement with the results
of a recent QCD sum rule analysis of the isospin
violating $<0\vert T(A^3_\mu A^8_\nu)\vert 0>$ correlator\cite{kmmb}.

If one considers allowing
an admixture of the singlet pseudoscalar current into the 
interpolating field (i.e, allowing $\alpha$ to deviate from
$-1/3$),
one has, of course, in addition to the $8_F\times 8_F\rightarrow 1_F$
reduced matrix element which governs $<0|P_1^0|\eta^\prime >$ and
$<0|P_0^8|\eta^\prime >$, the $1_F\times 1_F\rightarrow 1_F$ reduced
matrix element relevant to $<0|P_0^0|\eta^\prime >$.  One can then
certainly, in principle, find a value of $\alpha$ such that
$<0|P(\alpha )|\eta^\prime >=0$.  The problem is that, even to
do so at leading order in the quark masses, one would need to know
the ratio of the two reduced matrix elements above, and this
information is not available.  Moreover, even if it were, this
would not necessarily ensure that, for such a value of $\alpha$, the
couplings of the higher resonances other than the $\eta^\prime$ 
were small for the same value
of $\alpha$.

Since we do not know, {\it a priori}, how to choose a $\pi^0$
interpolating field (i.e., a value of $\alpha$) to remove even $\eta^\prime$
contamination, let alone possible contamination associated with
yet higher pseudoscalar resonances, it is necessary to look for
some sort of {\it post facto} indication of the absence of such
contributions.  One obvious way of doing so is to study the
extracted results for what is nominally the isospin violating
$\pi NN$ coupling, $\delta g$, as a function of $\alpha$.  If
one can find a region of $\alpha$ values for which the results
are not sensitive to $\alpha$, then one might argue this was
a signal that, for such values of $\alpha$, the effect of couplings
to all higher resonances is negligible.  In contrast, if one is
unable to find such a region of $\alpha$ values then, since the
various interpolating fields differ only in their couplings to
the excited pseudoscalar resonances, beginning with the $\eta^\prime$,
it is clear that, in general, there are large contaminations from
the higher pseudoscalar resonances and that, as a consequence,
one has no reliable way of choosing a particular interpolating
field for which such contaminations are small.  We will see,
unfortunately, that it is the latter situation which holds for
the 3-point function analysis.  Moreover, we will demonstate
that the variation of $\delta g$ with $\alpha$ is essentially
as large as the value of $\delta g$ extracted in the Meissner-Henley
analysis, and hence that no reliable estimate of $\delta g$ can
be made using the 3-point function method.

In order to demonstrate the claims of the last paragraph, it is
necessary to understand how the generalization from the
specific Meissner-Henley interpolating field choice, $\alpha =-1/3$,
to arbitrary $\alpha$ affects the sum rule for the isospin breaking
$\pi NN$ coupling.  It is straightforward to show that, in the
general case, the final sum rule (the analogue of Eq.~(24) of
Ref.\cite{mh97}) becomes
\begin{eqnarray}
\left[ - \frac{\delta g}{g_{\pi NN}} \right] \,
&=& - \frac{2}{3} \gamma \, +\, \frac{4}{3} \kappa (\alpha )
\, +\, 
\left ( \frac{\delta M_N}{M_N} \right ) \nonumber \\
&&\qquad \frac{(2\pi)^2 {m_\pi}^2 {f_\pi}^2 M^2}
{M^6 + \frac{1}{4} g_s ^2 <G^2> M^2 }
\left[ 2\left( \frac{m_d - m_u}{m_d + m_u}\right)\, +\, 
\gamma - 2  \frac{\delta M_N}{M_N} 
\frac{{M_N}^2}{M^2} \right ] ,
\label{mhmod}
\end{eqnarray}
where $M$ is the Borel mass, $<G^2>$ the gluon condensate,
$g_s$ the strong coupling constant, $\delta M_N$ the quark-mass-difference
contribution to the nucleon mass splitting,
and $\kappa (\alpha )$ the coefficient of
$P_u+P_d$ appearing in the interpolating field $P (\alpha )$ (so, for example,
for the Meissner-Henley choice $P_{\pi^0}$, $\kappa = \theta_0/\sqrt{3}$).
The terms in the second line of Eq.~(\ref{mhmod}) arise from evaluating
the isospin breaking difference of couplings of the Ioffe currents
to the neutron and proton states using the chiral odd sum rule for
the nucleon two-point functions\cite{mh97}.  The reason for the
appearance of $\kappa (\alpha )$ in Eq.~(\ref{mhmod}), i.e., the
dependence on only the light quark $I=0$ content of the interpolating
field, is that contributions of the strange quark component of the 
interpolating field to the OPE side of the sum rule vanish to the order 
considered in obtaining the sum rule.

From the expression for $P (\alpha )$ in Eq.~(\ref{alphafamily}),
it is evident that
\begin{equation}
\kappa (\alpha )= \alpha\epsilon (\alpha )\ .
\end{equation}
In order to complete our investigation we, therefore, require only
the expression for $\epsilon (\alpha )$, obtained in Ref.~\cite{kmcw}:
\begin{eqnarray}
\epsilon(\alpha )&=& -{\sqrt 3}\theta_0\Biggl[ 1
+\left(
{\frac{-(10+9\alpha )\mu_\pi +6(1+\alpha )\mu_K +(4+3\alpha )\mu_\eta}{3F^2}}
\right)\nonumber \\
&&\qquad -{\frac{32}{3F^2}}
(4+3\alpha )\left( \bar{m}_K^2-m_\pi^2 \right)\left( 3L_7^r+L_8^r\right)
+\left({\frac{3m_\eta^2 +m_\pi^2}{64\pi^2F^2}}\right)
\Biggl( 1+\nonumber \\
&&\qquad \left[ {\frac{m_\pi^2}{\bar{m}_K^2-m_\pi^2}}\right]\log\left(
{\frac{m_\pi^2}{\bar{m}_K^2}}\right)\Biggr)
+{\frac{(m_\eta^2-m_\pi^2)}{64\pi^2F^2}}\left( 1+\log (m_K^2/\mu^2)\right) 
\Biggr]\ , 
\label{epsalpha} 
\end{eqnarray}
where $\bar{m}_K^2$ is the average of the $K^+$ and $K^0$ squared
masses,
the chiral log terms $\mu_M$ are as defined above and the
$L_k^r$ are the usual renormalized fourth order LEC's, in the
notation of Gasser and Leutwyler\cite{gl85}.
In Eq.~(\ref{epsalpha}) the expression has been written so that the
$1$ occurring in the square brackets corresponds to the contribution
obtained at leading order in the chiral expansion, while the
remaining terms give the next-to-leading order corrections.
The above results reproduce the Meissner-Henley field choice for
$\alpha =-1/3$, if one keeps only the leading order contribution
in Eq.~(\ref{epsalpha}).

If we take the latest evaluation of the quark mass ratios from 
ChPT\cite{leutwyler96}, then we have
\begin{equation}
\theta_0 =\, (1.1\pm 0.2)\times 10^{-2}.
\end{equation}
(The upper end of the error bound would correspond to the evaluation
of $r=(m_d-m_u)/(m_d+m_u)$ obtained using $\eta\rightarrow 3\pi$, which
corresponds to large values for the violation of Dashen's theorem
advocated by a number of authors\cite{mk90,dhw93,bijnens93,dp96,det96,bp96}.)
For the Meissner-Henley choice $\alpha =-1/3$, we then find, for the
contribution to $-\delta g/g_{\pi NN}$ generated by the isoscalar
component of the $\pi^0$ interpolating field (required to remove
the $\eta$ contamination from the final result),
\begin{equation}
\left[ -\delta g/g_{\pi NN}\right]_{\kappa (-1/3)}\, =\, 
\frac{4}{3}\frac{\theta_0}{\sqrt{3}}[1+0.27] =\, 1.1\times 10^{-2}\ ,
\end{equation}
while, for comparison, for the choice $\alpha =-1$, which removes the
strange quark content from the interpolating field, we find
\begin{equation}
\left[ -\delta g/g_{\pi NN}\right]_{\kappa (-1)}\, =\, 
\frac{4}{3}\sqrt{3}\, \theta_0[1+0.21] =\, 3.1\times 10^{-2}\ .
\end{equation}
To understand the implication of these results, one should bear in mind
that the range of values for $-\delta g/g_{\pi NN}$ extracted by
Meissner and Henley was $(1.7$--$3.0)\times 10^{-2}$.  (The
range reflects the full range of uncertainties in all of the input
parameters, $\gamma$, $\theta_0$ and $\delta M_N$.)
We thus find
that the corrections required to remove the $\eta$ contamination
are large on the scale of the result obtained.  Moreover, since
$\kappa (\alpha )=\alpha \epsilon (\alpha )$, and the results of
Ref.~\cite{kmcw} show $\epsilon (\alpha )$ to be slowly varying
with $\alpha$, and greater than $1$ over a wide range of $\alpha$
values, we see also that the results for what is nominally the
isospin breaking coupling is very sensitive to $\alpha$, varying,
for example, by an amount as large as the maximum value
quoted by Meissner and Henley over the range between $\alpha = 0$
and $\alpha = -1$.  We thus conclude, in light of the discussion
above, that the 3-point function evaluation of the isospin breaking
coupling is plagued by unknown higher resonance contamination,
and as such cannot provide a reliable estimate of this quantity.

\section{Conclusions}
We have shown that the 3-point function method for the treatment
of both the isospin conserving and isospin violating $\pi NN$
couplings is plagued by problems with higher resonance contamination.
In the course of this investigation we have also seen how information
from the finite energy sum rules for the light quark masses, chiral
perturbation theory, and sum rules for the chiral LEC's can
sometimes be profitably employed to elucidate the physical content
of other sum rules treatments. 

\acknowledgements
The author would like to acknowledge a number of useful discussions with
M. Birse, and also with E. Henley and T. Meissner on the contents of 
Ref.\cite{mh97}.  The ongoing support of the Natural Sciences and
Engineering Research Council of Canada, and the hospitality of the
Special Research Centre for the Subatomic Structure of Matter at the
University of Adelaide, where this work was performed, are also
gratefully acknowledged.

\end{document}